\documentclass[12pt]{article}
\usepackage{epsfig}
\usepackage{amsmath}
\newcommand{\aem}{\alpha_{\mathrm{em}}}
\renewcommand{\d}{\mathrm{d}}
\newcommand{\X}{\mathbf{X}}
\newcommand{\e}{\mathrm{e}}
\newcommand{\f}{\mathrm{f}}
\newcommand{\g}{\mathrm{g}}
\newcommand{\p}{\mathrm{p}}
\newcommand{\q}{\mathrm{q}}
\newcommand{\fbar}{\mathrm{\overline{f}}}
\newcommand{\qbar}{\mathrm{\overline{q}}}

\newcommand{\kT}{k_{\perp}}
\newcommand{\pT}{p_{\perp}}

\newcommand{\ET}{E_{\perp}}
\newcommand{\shat}{\hat{s}}

\newcommand{\gast}{\gamma^*}%

\def\Journal#1#2#3#4{{#1}{\bf #2} (#4) #3}

\def\NPB{{\rm Nucl. Phys.~}{\bf B}}
\def\PLB{{\rm Phys. Lett.~}{\bf B}}
\def\JournalPLB#1#2#3{{\rm Phys. Lett.~}{\bf {#1}B} (#3) #2}

\def\PRD{{\rm Phys. Rev.~}{\bf D}}
\def\ZPC{{\rm Z.~Phys.~}{\bf C}}
\def\ZP{\rm Z.~Phys.~}

\def\CPC{\rm Computer~Phys.~Commun.~}
\def\PRP{\rm Phys. Rep.~}
\def\PRV{\rm Phys. Rev.~}
\def\EPJC{{\rm Eur. Phys.~J.~}{\bf C}}
\newlength{\abstwidth}
\begin{document}
 
\sloppy
 
\pagestyle{empty}

\begin{flushright}
LU TP 99--16 \\
hep-ph/9907299\\
July 1999
\end{flushright}
 
\vspace{\fill}

\begin{center}  \begin{Large} \begin{bf}
Jet Production by Virtual Photons%
\footnote{To appear in the Proceedings of the %
International Conference on %
the Structure and Interactions %
of the Photon; Photon 99, 23-27 May 1999, %
Freiburg im Breisgau, Germany.}\\[10mm]
  \end{bf}  \end{Large}
  \vspace*{5mm}
  \begin{large}
C. Friberg\\[2mm]
  \end{large}
 Department of Theoretical Physics, Lund University,\\ 
     Helgonav\"agen 5, S-223 62 Lund, Sweden\\ 
     christer@thep.lu.se
\end{center}

\vspace{\fill}

\begin{center}
{\bf Abstract}\\[2ex]
\begin{minipage}{\abstwidth}
The production of jets is studied in collisions of virtual photons,
$\gamma^* \mathrm{p}$ and $\gamma^*\gamma^*$, specifically for applications 
at HERA and LEP2. Photon flux factors are convoluted with matrix elements 
involving either direct or resolved photons and, for the latter, with parton
distributions of the photon. Special emphasis is put on the range of 
uncertainty in the modeling of the resolved component. The resulting model 
is compared with existing data.
\end{minipage}
\end{center}

\vspace{\fill}

\clearpage
\pagestyle{plain}
\setcounter{page}{1} 

\section{Introduction}

The photon is a complicated object to describe. In the DIS region, i.e. 
when it is very virtual, it can be considered as devoid of any internal
structure, at least to first approximation. In the other extreme, the 
total cross section for real photons is dominated by the resolved 
component of the wave function, where the photon has fluctuated 
into a $\q\qbar$ state. The nature of this resolved component is
still not well understood, especially not the way in which it dies 
out with increasing photon virtuality. This dampening is likely not 
to be a simple function of virtuality, but to depend on the physics
observable being studied, i.e. on the combination of subprocesses
singled out.  

Since our current understanding of QCD does not allow complete 
predictability, one sensible approach is to base ourselves on
QCD-motivated models, where a plausible range of uncertainty can be 
explored. Hopefully comparisons with data may then help constrain
the correct behaviour. The ultimate goal therefore clearly is to
have a testable model for all aspects of the physics of $\gast\p$ 
and $\gast\gast$ collisions. As a stepping stone towards 
constructing such a framework, in this paper we explore the physics 
associated with the production of `high-$\pT$' jets in the collision. 
That is, we here avoid the processes that only produce activity along 
the $\gast\p$ or $\gast\gast$ collision axis. For resolved photons this
corresponds to the `soft' or `low-$\pT$' events of the hadronic 
physics analogy, for direct ones to the lowest-order DIS process
$\gast\q \to \q$. 

The processes that we will study here instead can be exemplified by
$\gast\gast \to \q\qbar$ (direct), $\gast \g \to \q\qbar$ 
(single-resolved for $\gast\gast$, direct for $\gast\p$) and 
$\g\g \to \q\qbar$ (double-resolved for $\gast\gast$, 
(single-)resolved for $\gast\p$), where the gluons come from the 
parton content of a resolved virtual photon or from the proton.
Note that these are multi-scale processes, at least involving the
virtuality $Q_i^2$ of either photon ($i=1,2$) and the $\pT^2$ of 
the hard subprocess. 
For a resolved photon, the relative transverse momentum $\kT$ of 
the initial $\gast \to \q\qbar$ branching provides a further scale,
at least in our framework. 
 
Almost real photons allow long-lived $\gast \to \q\qbar$ fluctuations, 
that then take on the properties of non-perturbative hadronic states,
specifically of vector mesons such as the $\rho^0$.  It is therefore
that an effective description in terms of parton distributions becomes
necessary. Hence the resolved component of the photon, as opposed to
the direct one.
That such a subdivision is more than a technical construct is
excellently illustrated by the $x_{\gamma}^{\mathrm{obs}}$ plots from
HERA \cite{xgobs}.

The resolved photon can be further subdivided into low-virtuality
fluctuations, which then are of a nonperturbative character and can be
represented by a set of vector mesons, and high-virtuality ones that
are describable by perturbative $\gast \to \q\qbar$ branchings.  The
former is called the VMD (vector meson dominance) component and the
latter the anomalous one. The parton distributions of the VMD
component are unknown from first principles, and thus have to be based
on reasonable ans\"atze, while the anomalous ones are perturbatively
predictable. 

The traditional tool for handling such complex issues is the Monte Carlo
approach. Our starting point is the model for real photons 
\cite{sasevt} and the parton distribution parameterizations of real
and virtual photons \cite{saspdf} already present in the {\sc Pythia}
\cite {pythia} generator. Several further additions and modifications
have been made to model virtual photons, as will be described in the
following~\cite{paper}. 

\section{The Model}
\label{model}

The cross sections for the processes $\e\p\rightarrow\e\X$ and 
$\e\e\rightarrow\e\e\X$ can be written as the 
convolutions~\cite{weiz,will,EPA}
\begin{equation}
\d\sigma(\e\p\rightarrow\e\X)=\hspace{-3mm}
\sum_{\xi=\mathrm{T,L}}
\iint \d y \, \d Q^2 
\;f_{\gamma/\e}^{\xi}(y,Q^2) 
\;\d\sigma(\gast_{\xi}\p\rightarrow\X)
\label{PYHERAEPA}
\end{equation}
and 
\begin{equation}
\d\sigma(\e\e\rightarrow\e\e\X)=\hspace{-5mm}
\sum_{\xi_1,\xi_2=\mathrm{T,L}}
\iiiint \!\!\!\d y_1  \d Q_1^2  \d y_2 \, \d Q_2^2 \,
f_{\gamma/\e}^{\xi_1}(y_1,Q_1^2) f_{\gamma/\e}^{\xi_2}(y_2,Q_2^2)
\d\sigma(\gast_{\xi_1}\gast_{\xi_2}\rightarrow\X).
\label{PYLEPEPA}
\end{equation}
The flux of photons $f(y,Q^2)$ (see $y$ definition below) from the lepton
is factorized from the subprocess cross sections involving the virtual 
photon, $\gast\p\rightarrow\X$ and $\gast\gast\rightarrow\X$. The sum is
over the transverse and longitudinal photon polarizations. For $\e\p$ 
events, this factorized ansatz is perfectly general, so long 
as azimuthal distributions in the final state are not studied in detail.
In $\e^+\e^-$ events, it is not a good approximation when the 
virtualities $Q_1^2$ and $Q_2^2$ of both photons become of the order of 
the squared invariant mass $W^2$ of the colliding photons 
\cite{GS}. 

When $Q^2/W^2$ is small, one can derive~\cite{gammaflux,EPA,GS}
\begin{eqnarray}
f_{\gamma/l}^{\mathrm{T}}(y,Q^2) & = & \frac{\aem}{2\pi} 
\left( \frac{(1+(1-y)^2}{y} \frac{1}{Q^2}-\frac{2m_{l}^2y}{Q^4} \right) ~,\\
f_{\gamma/l}^{\mathrm{L}}(y,Q^2) & = & \frac{\aem}{2\pi} 
\frac{2(1-y)}{y} \frac{1}{Q^2} ~.
\label{LLogflux}
\end{eqnarray}
The $y$ variable is defined as the lightcone fraction the photon takes 
of the incoming lepton momentum. The lepton scattering angle $\theta_i$ 
is related to $Q_i^2$, where the kinematical limits 
on $Q_i^2$ are, unless experimental conditions reduce the $\theta$ range, 
$Q^2_{i,\mathrm{min}} \approx \frac{y^2}{1 - y} m_{\e}^2$ and
$Q^2_{i,\mathrm{max}} \approx (1 - y) s$. 

Within the allowed region, the phase space is Monte Carlo sampled according to
$(\d Q^2/Q^2) \, (\d y / y) \, \d \varphi$, with the remaining flux factor 
combined with the cross section factors to give the event weight used for 
eventual acceptance or rejection.

The hard-scattering processes are classified according to whether 
one or both photons are resolved. For the direct process 
$\gast_{\xi_i} \gast_{\xi_i} \to \f\fbar$, $\f$ some fermion, the cross 
sections for transverse and longitudinal photons are used~\cite{Baier,paper}.
Remember that the cross section for a longitudinal photon vanishes as 
$Q_i^2$ in the limit $Q_i^2 \to 0$. 

For a resolved photon, the photon virtuality scale is included in the 
arguments of the parton distribution but, in the spirit of the parton 
model, the virtuality of the parton inside the photon is not included 
in the matrix elements. Neither is the possibility of the partons being
in longitudinally polarized photons (see below, however). The same 
subprocess cross sections can therefore be used for direct $\gast \p$ 
processes and for single-resolved $\gast\gast$ ones. Both transverse 
and longitudinal photons are considered for the QCD Compton 
$\gast \q \to \g \q$ and boson--gluon fusion $\gast \g \to \q \qbar$
processes~\cite{siggap,paper}. The matrix elements are then convoluted 
with parton distributions. 

Finally we come to resolved processes in $\gast\p$ and doubly-resolved 
ones in $\gast\gast$. There are six basic QCD cross sections, 
$\q\q' \to \q\q'$, $\q\qbar \to \q'\qbar'$, $\q\qbar \to \g\g$,
$\q\g \to \q\g$, $\g\g \to \g\g$ and $\g\g \to \q\qbar$. The same 
subprocess cross sections as those known from $\p\p$ 
physics~\cite{sigpp} can therefore be used. Again, a convolution with 
parton distributions is necessary. 

One major element of model dependence enters via the choice of parton
distributions for a resolved virtual photon. These distributions
contain a hadronic component that is not perturbatively calculable. It
is therefore necessary to parameterize the solution with input from
experimental data, which mainly is available for (almost) real
photons. In the following we will use the SaS distributions
\cite{saspdf}, which are the ones best suited for our
formalism. Another set of distributions is provided by GRS
\cite{grspdf}, while a simpler recipe for suppression factors relative
to real photons has been proposed by DG \cite{dgpdf}.

The SaS distributions for a real photon can be written as
\begin{equation}
f_a^{\gamma}(x,\mu^2) =
\sum_V \frac{4\pi\aem}{f_V^2} f_a^{\gamma,V}(x,\mu^2; Q_0^2)
+ \frac{\aem}{2\pi} \, \sum_{\q} 2 e_{\q}^2 \,
\int_{Q_0^2}^{\mu^2} \frac{{\d} k^2}{k^2} \,
f_a^{\gamma,\q\qbar}(x,\mu^2;k^2) ~.
\label{decomp}
\end{equation}
Here the sum is over a set of vector mesons
$V = \rho^0, \omega, \phi, \mathrm{J}/\psi$ according to a 
vector-meson-dominance ansatz for low-virtuality fluctuations of
the photon, with experimentally determined couplings $4\pi\aem/f_V^2$.
The higher-virtuality, perturbative, fluctuations are 
represented by an integral over the virtuality $k^2$ and a sum over
quark species. We will refer to the first part as the VMD one and 
the second as the anomalous one.  

From the above ansatz, the extension to a virtual photon is given by
the introduction of a dipole dampening factor for each component,
\begin{eqnarray}
f_a^{\gast}(x,\mu^2,Q^2)
& = & \sum_V \frac{4\pi\aem}{f_V^2} \left(
\frac{m_V^2}{m_V^2 + Q^2} \right)^2 \,
f_a^{\gamma,V}(x,\mu^2;\tilde{Q}_0^2)
\nonumber \\
& + & \frac{\aem}{2\pi} \, \sum_{\q} 2 e_{\q}^2 \,
\int_{Q_0^2}^{\mu^2} \frac{{\d} k^2}{k^2} \, \left(
\frac{k^2}{k^2 + Q^2} \right)^2 \, f_a^{\gamma,\q\qbar}(x,\mu^2;k^2)~.
\label{decompvirt}
\end{eqnarray}
Thus, with increasing $Q^2$, the VMD components die away faster than
the anomalous ones, and within the latter the low-$k^2$ ones faster
than the high-$k^2$ ones.   

Since the probed real photon is purely transverse, the above ansatz
does not address the issue of parton distributions of the longitudinal
virtual photons. One could imagine an ansatz based on longitudinally
polarized vector mesons, and branchings $\gast_{\mathrm{L}} \to
\q\qbar$, but currently no parameterization exists along these
lines. We will therefore content ourselves by exploring 
alternatives based on applying a simple
multiplicative factor $R$ to the results obtained for a resolved
transverse photon. As usual, processes involving longitudinal photons
should vanish in the limit $Q^2 \to 0$. To study two extremes, the 
region with a linear rise in $Q^2$ is defined either by $Q^2<\mu^2$
or by $Q^2<m_{\rho}^2$, where the former represents the perturbative
and the latter some non--perturbative scale. Also the high-$Q^2$ limit 
is not well constrained; we will compare two different alternatives, 
one with an asymptotic fall-off like $1/Q^2$ and another which 
approaches a constant ratio, both with respect to the transverse resolved 
photon. (Since we put $f_a^{\gast}(x,\mu^2,Q^2) = 0$ for $Q^2>\mu^2$, 
the $R$ value will actually not be used for large $Q^2$, so the 
choice is not so crucial.) We therefore study the alternative 
ans\"atze
\begin{eqnarray}
R_1(y,Q^2,\mu^2) & = & 1 + a \frac{4 \mu^2 Q^2}{(\mu^2 + Q^2)^2}
\frac{f_{\gamma/l}^{\mathrm{L}}(y,Q^2)}
{f_{\gamma/l}^{\mathrm{T}}(y,Q^2)}~,\label{Rfact1}\\
R_2(y,Q^2,\mu^2) & = & 1 + a \frac{4 Q^2}{(\mu^2 + Q^2)}
\frac{f_{\gamma/l}^{\mathrm{L}}(y,Q^2)}
{f_{\gamma/l}^{\mathrm{T}}(y,Q^2)}~,\label{Rfact2}\\
R_3(y,Q^2,\mu^2) & = & 1 + a \frac{4 Q^2}{(m_{\rho}^2 + Q^2)}
\frac{f_{\gamma/l}^{\mathrm{L}}(y,Q^2)}
{f_{\gamma/l}^{\mathrm{T}}(y,Q^2)}\label{Rfact3}
\end{eqnarray}
with $a=1$ as main contrast to the default $a=0$. The $y$ dependence 
compensates for the difference in photon flux between transverse
and longitudinal photons. 

Another ambiguity is the choice of $\mu^2$ scale in parton distributions.
Based on various considerations, we compare three different alternatives: 
\begin{equation}
\mu_1^2 =  \pT^2 \frac{\shat+ Q_1^2 + Q_2^2}{\shat} ~,~~~~
\mu_2^2 =  \pT^2 + Q_1^2 + Q_2^2 ~,~~~~
\mu_3^2 =  2 \mu_1^2~.
\label{mu}
\end{equation}  
Only the second alternative ensures $f_a^{\gast}(x,\mu^2,Q^2) > 0$ for
arbitrarily large $Q^2$; in all other alternatives the resolved
contribution (at fixed $\pT$) vanish above some $Q^2$ scale. The last 
alternative exploits the well-known freedom of including some multiplicative
factor in any (leading-order) scale choice.
When nothing is mentioned explicitly below, the choice $\mu_1^2$ is used.

The issues discussed above are the main ones that distinguish the
description of processes involving virtual photons from those 
induced by real photons or by hadrons in general. In common is 
the need to consider the buildup of more complicated partonic
configurations from the lowest-order `skeletons' defined above, 
(\textit{i}) by parton showers, (\textit{ii}) by multiple 
parton--parton interactions and beam remnants, where applicable,
and (\textit{iii}) by the subsequent transformation of these partons 
into the observable hadrons. The latter, hadronization stage can be 
described by the standard string fragmentation framework 
\cite{AGIS}, followed by the decays of unstable primary hadrons, 
and is not further discussed here. The parton shower, 
multiple-interaction and beam-remnant aspects are discussed 
elsewhere~\cite{paper}.

\section{Comparisons with Data}
\label{migration}

In this section the model is compared with data. We will not make a detailed 
analysis of experimental results but use it to point out model dependences
and to constrain some model parameters. 

\subsection{Inclusive Jet Cross Sections}

\begin{figure*} [!htb]
   \begin{center}    
     \mbox{\psfig{figure=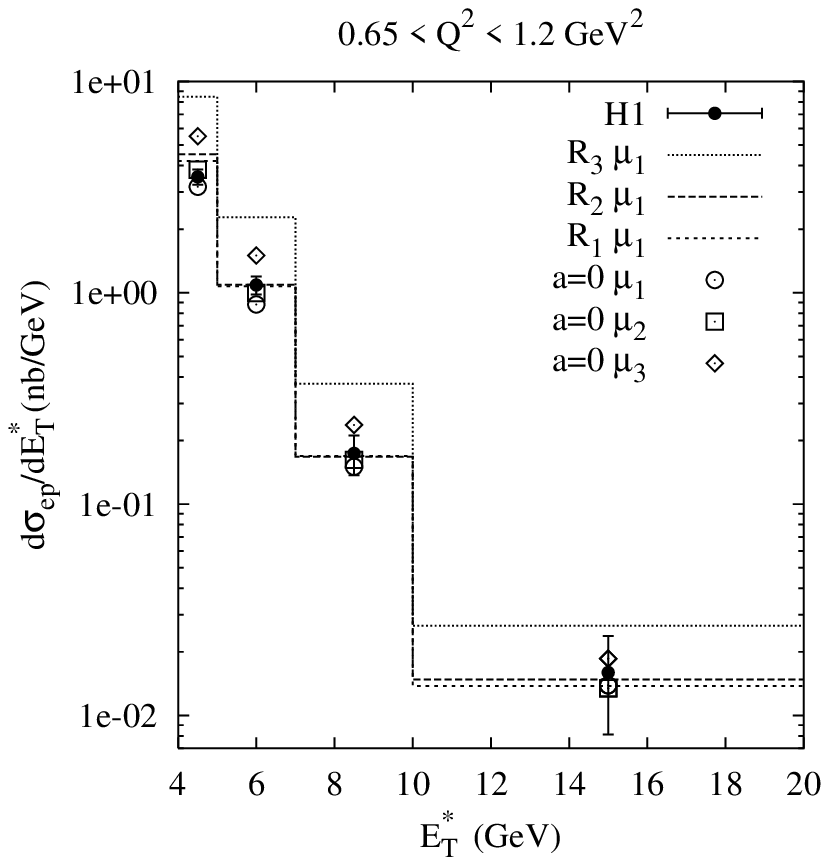,width=78mm}\hspace{-0.5cm}
	   \psfig{figure=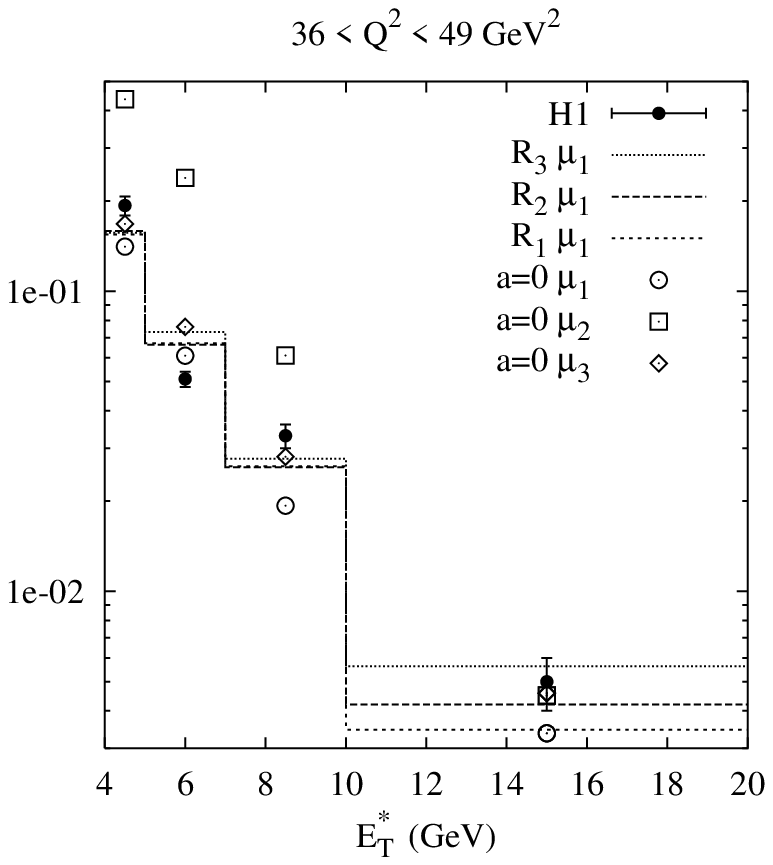,width=78mm}}
   \end{center}
\caption{ The differential jet cross section $\d\sigma_{\e\p}/\d E^*_{\perp}$ 
for jets with $-2.5<\eta^*<-0.5$ and $0.3<y<0.6$. $a=0$ indicates that only 
transversely polarized resolved photons are considered.}
\label{fig:ETL}
\end{figure*}

Inclusive $\e\p$ jet cross sections have been measured by the H1 collaboration 
\cite{lowQ2H1} in the kinematical range $0<Q^2<49~\mathrm{GeV}^2$ and 
$0.3<y<0.6$. For $\d\sigma_{\e\p}/\d E^*_{\perp}$ data is available in nine 
different $Q^2$ bins; two of them are shown here, Fig.~\ref{fig:ETL},
with similar results for the intermediate bins. The plots were produced 
with the HzTool~\cite{hz} package. The $E^*_{\perp}$ and $\eta^*$ are 
calculated in the $\gast\p$ centre of mass frame where the incident proton 
direction corresponds to positive $\eta^*$. The SaS~1D parton distribution 
together with a few different $\mu_i$ scales are used to model the resolved 
photon component. 

In the highest $Q^2$ bin the direct component gives the dominant contribution.
However, the resolved component is not negligible. All the scales 
$\mu_i$  depend on the photon virtuality. This gives a larger resolved 
component in this region as compared to the conventional choice, 
$\mu=p_{\perp}$. In the low $Q^2$ bin the $\mu_1$ and $\mu_2$ scales do not 
differ much from $p_{\perp}$ and the cross sections are in nice agreement with 
data. The cross section with the $\mu_3$ scale overshoots the data in this 
region. 

Changing the photon parton distribution from SaS~1D to SaS~2D will give a 
slightly lower result for the low-$Q^2$ bins.
A comparison has been made with the GRS~LO~\cite{grspdf} parton 
distribution~\cite{paper} and, in its region of validity, differences are 
small. One could imagine larger differences for $\gast$ parton distributions 
that from the onset are more different. Using a parton distribution for 
a real photon cannot describe the $Q^2$ dampening in the distributions 
shown in this section. 

Using CTEQ~3L instead of GRV~94~LO (which is default in {\sc Pythia}) as 
the proton parton distribution reduces the result in some $\ET^*$ bins by 
half. The GRV~94~HO parton distribution give a slightly 
lower result (as compared to GRV~94~LO). The differences mainly come from 
the gluon distributions, that are not yet so well constrained from data. 
In the modeling of the parton distributions, it is a deceptive accident that 
the more well-known proton parton distribution gives a larger uncertainty 
than the photon one. It offers a simple example that also phenomenology of 
other areas may directly influence the interpretation of photon data. 

The OPAL collaboration has measured inclusive one--jet and two--jet cross 
sections in the range $|\eta^\mathrm{jet}|<1$ and requiring 
$\ET^\mathrm{jet}$ to be larger than 3~GeV~\cite{OPAL}. The centre of mass 
energies were 130 and 136~GeV. The inclusive jet cross sections as a 
function of $\ET^\mathrm{jet}$ or $\eta^\mathrm{jet}$ are compared with 
data~\cite{paper}, with events generated at $\sqrt{s_{\e\e}}=133$~GeV. 

At low $\ET^\mathrm{jet}$ the double--resolved events are dominating 
and at larger $\ET^\mathrm{jet}$ it is the direct processes since more 
energy goes into the hard scattering in the latter case. For 
single--resolved events, the SaS~1D VMD component dies out much quicker 
with increasing $\ET^\mathrm{jet}$ than the SaS~2D one which is comparable 
with the direct--anomalous events at high $\ET^\mathrm{jet}$. For both 
cases, at high $\ET^\mathrm{jet}$, the direct--anomalous components give 
the same order of magnitude contribution to the cross section as the 
double--resolved events. The biggest 
difference between the two parton distributions can be seen at low 
$\ET^\mathrm{jet}$ and for the $|\eta^\mathrm{jet}|$ distributions, where the 
double--resolved events dominate; it is a reflection of the difference in
normalization among the contributions. For the SaS~2D case, this kinematical 
region makes the VMD component more important than the anomalous one; as a 
consequence multiple interactions play an important role. The double--resolved 
contribution for SaS~2D without multiple interaction is reduced by half. 
Clearly, for the SaS~1D case the opposite is true: the 
importance of the components are reversed. In the region of high 
$\ET^\mathrm{jet}$, where the direct events dominate, the model is 
under\-shooting data. On the other hand, there is nice agreement with data for 
the $|\eta^\mathrm{jet}|$ distribution when using SaS~2D.

\subsection{Forward Jet Cross Sections}
\label{secfwdep}

Jet cross sections as a function of Bjorken-$x$, $x_{\mathrm{Bj}}$, for 
forward jet production (in the proton direction) have been measured at 
HERA~\cite{fwdjet}. The objective is to probe the dynamics of the QCD cascade 
at small $x_{\mathrm{Bj}}$. The forward jet is restricted in polar angle 
w.r.t. the proton and the transverse momenta $p_{\perp}^\mathrm{jet}$ should 
be of the same order as the virtuality of the photon, suppressing an 
evolution in transverse momenta. If the jet has a large energy fraction of 
the proton there will be a big difference in $x$ between the jet and the 
photon vertex; $x_{\mathrm{Bj}} \ll x_{\mathrm{jet}}$, allowing an evolution 
in $x$. The above restrictions will not eliminate the possibility of having a 
resolved photon, although the large $Q^2$ values are not in favour of it. 

\begin{figure*} [!htb]
   \begin{center}     
   \mbox{\psfig{figure=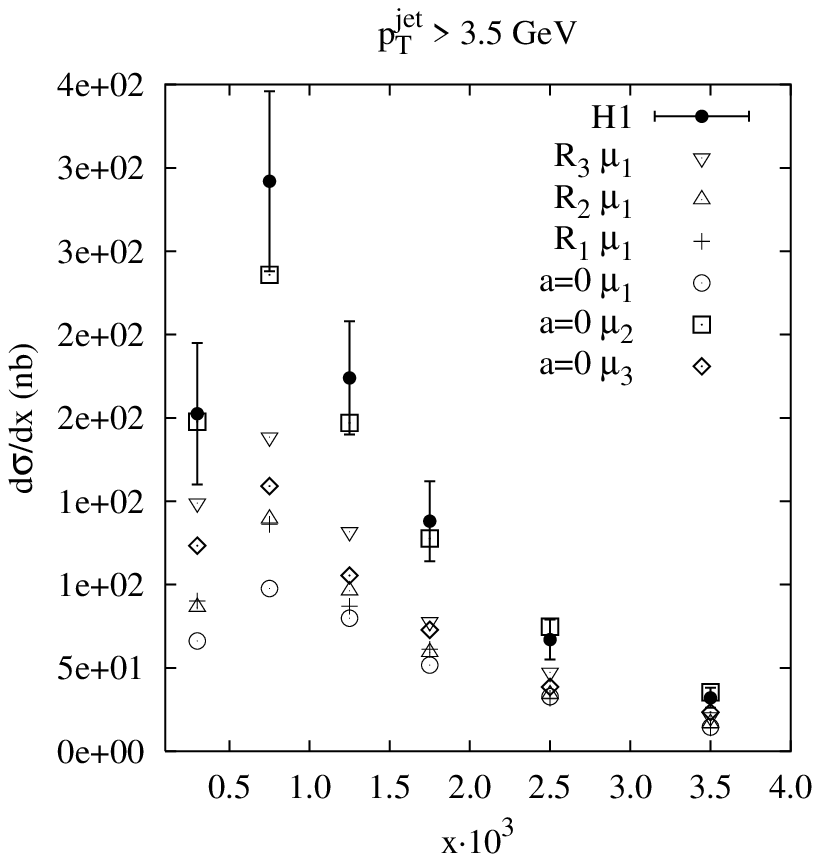,width=78mm}\hspace{-0.5cm}
	\psfig{figure=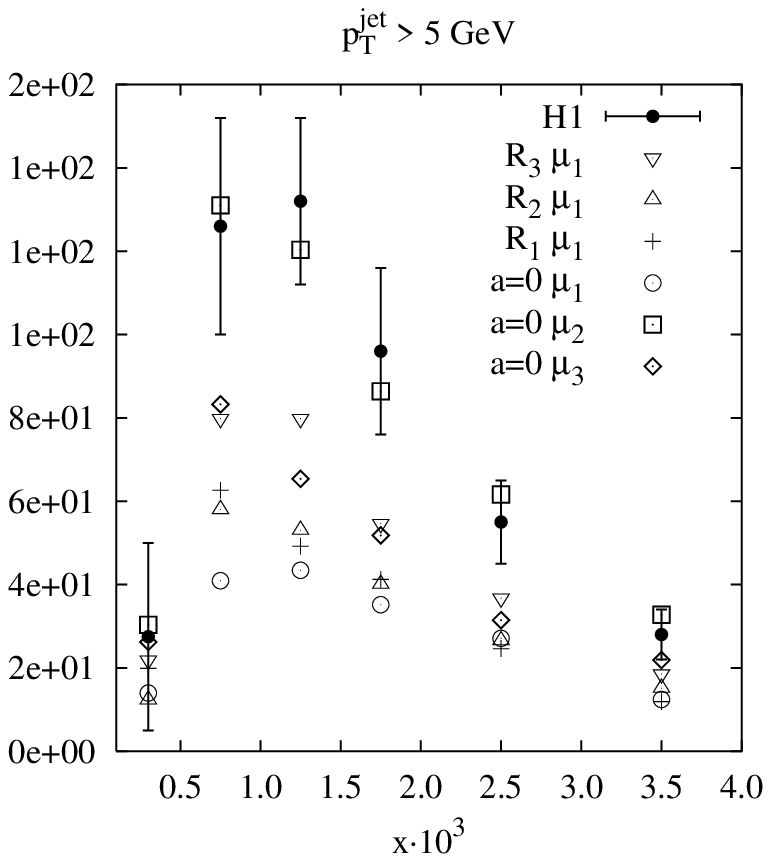,width=78mm}}
   \end{center}
\caption{Forward jet cross section as a function of $x$. The results
with three different alternatives of longitudinal resolved photons $R_i$ 
are compared with purely transverse ones, $a=0$, and data from H1.}
\label{fig:fwdL}
\end{figure*}

The HzTool routines were used to obtain the results in 
Fig.~\ref{fig:fwdL}. A larger forward jet cross section is obtained with a 
stronger $Q^2$ dependence for the hard scale, with $\mu_2^2=p_{\perp}^2+Q^2$ 
in best agreement with data~\cite{LeifHannes}. The 
choice of scale does not only affect the resolved photon contribution but also 
the direct photon, arising from the scale dependence in the proton parton 
distribution. The rather large $Q^2$ values, 
$Q^2 \simeq (p_{\perp}^\mathrm{jet})^2$, suppresses VMD photons and favours the 
SaS~1D distribution which is the one used here, though the difference is small.

Note that the $\mu_3$ scale undershoots the forward jet cross section 
data and overshoots the inclusive jet distributions at low $Q^2$, 
so it is not a real alternative. As a further check, with more data 
accumulated and analysed, the $(p_{\perp}^\mathrm{jet})^2/Q^2$ interval 
could be split into several subranges which hopefully would help to 
discriminate between scale choices. 

With the experience of forward jets at HERA, we suggest a similar study at
LEP. The optimal kinematical and forward jet constraints have to be set by
each collaboration itself; we will only estimate the order of magnitude 
for the cross section and point out uncertainties in the model.

Comparing with forward jets at HERA, one of the leptons will play the role 
of the proton. Some of the constraints can be taken over directly, for 
example, $x_\mathrm{jet}=E_\mathrm{jet} / E_\e > 0.035$ and 
$0.5 < (\pT^\mathrm{jet})^2 / Q^2 < 2$. To fulfill the jet selection one of 
the leptons has to be tagged in order to know the virtuality of the photon. 
To obtain a reasonable number of events the other lepton is not tagged. 
With a centre of mass energy of 200~GeV, the smallest accessible 
$x_\mathrm{Bj}=\frac{Q^2}{y s}$ is around $10^{-4}$, where $Q^2$ and $y$ is 
calculated from the tagged electron, omitting the virtuality of the other 
photon. In a more sophisticated treatment also double--tagged events are 
analyzed; then one of the photons plays the role of a proton and the forward 
jet should be defined with respect to one of the photons. 

As for the case at HERA, the $\mu_2$ scale gives the 
largest forward jet cross section, about twice as large as with the $\mu_1$ 
scale~\cite{paper}. Most of the differences arise from the 
double--resolved events. Double--resolved and single--resolved events, where 
the resolved photon give rise to the forward jet, dominate the forward jet 
cross section. At low $x$, for the $\mu_2$ scale, the double--resolved 
contribution is close to an order of magnitude larger than the direct one. 
For the $\mu_1$ scale it is about a factor of four. As for the case at HERA, 
the rise of the forward jet cross section at small $x$ is dominated by 
resolved photons. A study like this at LEP could be an important cross 
check for the understanding of resolved photons and that of small-$x$ 
dynamics. 

\subsection{Importance of longitudinal resolved photons}

In this section we will study the importance of longitudinal resolved 
photons. A sensible $Q^2$--dependent scale choice, $\mu_1$, together with 
the SaS~1D distribution will be used throughout. 

With $a=1$ the different alternatives are shown in Fig.~\ref{fig:ETL} 
for the $\d\sigma_{\e \p}/\d E_{\perp}^*$ distributions together 
with the result from pure transverse photons, i.e. $a=0$. The importance 
of the resolved contributions decreases with increasing $Q^2$ which makes 
the asymptotic behaviour less crucial. 
The onset of longitudinal photons governed by the $R_1$ and $R_2$ 
alternatives are favoured whereas the $R_3$ one overshoots data in the 
context of the other model choices made here. 

In Fig.~\ref{fig:fwdL} the same alternatives are shown for the forward jet 
cross sections. With this scale choice, $\mu_1$, none of the longitudinal 
resolved components (together with the direct contribution) are sufficient 
to describe the forward jet cross section. The resolved contribution with 
$R_3$ is about the same as the one obtained with the scale 
$\mu_2^2=p_{\perp}^2+Q^2$ (without longitudinal contribution); 
the difference in the total results originates from the difference in the
direct contributions. With $R_1$ and $a=1$, the $\mu_2$ scale (not shown) 
overshoots the data.
The above study indicates, as expected, that longitudinal resolved photons
are important for detailed descriptions of various distributions. It cannot
by itself explain the forward jet cross section, but may give a significant 
contribution. Combined with other effects, for example, different scale 
choice, parton distributions and underlying events, it could give a 
reasonable description. The model(s) so far does not take into account the 
difference in $x$ distribution or the $k^2$ scale (of the 
$\gast \rightarrow \q\qbar$ fluctuations) between transverse and longitudinal 
photons. As long as the distributions under study allow a large interval in 
$x$ the average description may be reasonable. In a more sophisticated 
treatment these aspects have to be considered in more detail. 

\section{Summary and Outlook}

The plan here is to have a complete description of the main physics aspects 
in $\gamma\p$ and $\gamma\gamma$ collisions, which will allow important cross 
checks to test universality of certain model assumptions. As a step forward, 
we have here concentrated on those that are of importance for the 
production of jets by virtual photons, and are absent in the real-photon case. 
While we believe in the basic machinery developed and presented here, we have 
to acknowledge the many unknowns --- scale choices, parton distribution sets 
(also those of the proton), longitudinal contributions, underlying events, 
etc. --- that all give non-negligible effects. To make a simultaneous  
detailed tuning of all these aspects was not the aim here, but rather to 
point out model dependences that arise from a virtual photon. 

When $Q^2$ is not small, naively only the direct component needs to be 
treated, but in practice a rather large contribution arises from resolved 
photons. For example, for high-$Q^2$ studies like forward jet cross sections, 
Fig.~\ref{fig:fwdL}, or inclusive differential jet cross 
sections, Fig.~\ref{fig:ETL}. Resolved longitudinal 
photons are poorly understood and the model presented here can be used to 
estimate their importance and get a reasonable global description. 
Longitudinal effects are in most cases small but of importance for 
fine--tuning. 

The inclusive $\gast\gast$ one-jet and two-jet cross sections are well 
described except for the high $\ET^\mathrm{jet}$ region of the 
$\ET^\mathrm{jet}$ distribution~\cite{paper}. In this region, the direct 
events are dominating. Currently, owing to the lesser flexibility in the 
modeling of the direct component, we do not see any simple way to improve 
the model. The factorized ansatz made for the photon flux is expected to be 
valid in this kinematical range; interference terms are suppressed by 
$Q_1^2 Q_2^2/W_{\gast\gast}^2$. However, differences in the application of 
the cone jet algorithm may affect the results. 

The forward jet cross section presented by H1~\cite{fwdjet} is well described
by an ordinary parton shower prescription including the possibility of having
resolved photons. The criteria that the $\pT^\mathrm{jet}$ should be of
the same order as $Q^2$, makes the scale choice crucial and $\mu_2^2=\pT^2+Q^2$
is favoured by data, as concluded in~\cite{LeifHannes}. With this experience 
we predict the forward jet cross section to be obtained at LEP~\cite{paper}. 
With more data accumulated and analyzed, the $(p_{\perp}^\mathrm{jet})^2/Q^2$ 
interval could be split into several subranges, which hopefully would help to 
discriminate between different scale choices.

Multiple interactions for the anomalous component are not yet included, and is
not expected to be of same importance as in the VMD case. However, for 
low $k^2$ fluctuations it may be important, especially for SaS~1D, and need 
to be investigated. 

After this study of jet production by virtual photons it is natural to connect 
it together with low--$p_{\perp}$ events. Clearly, a smooth transition from 
perturbative to non--perturbative physics is required. Further studies are 
needed and will be presented in a future publication. 

\subsubsection*{Acknowledgements}

We acknowledge helpful conversations with, among others, Jon Butterworth, 
Jiri Ch\'yla, Gerhard Schuler, Hannes Jung, Leif J\"onsson, Ralph Engel and 
Tancredi Carli.

\end{document}